# Imaging ultrafast carrier transport in nanoscale devices using femtosecond photocurrent microscopy

B. H. Son, J. K. Park, J. T. Hong, J. Y. Park, S. Lee, and Y. H. Ahn*

*Department of Physics and Department of Energy Systems Research, Ajou University, Suwon 443-749, Korea*

One-dimensional nanoscale devices, such as semiconductor nanowires (NWs) and single-walled carbon nanotubes (SWNTs), have been intensively investigated because of their potential application of future high-speed electronic, optoelectronic, and sensing devices[1-3]. To overcome current limitations on the speed of contemporary devices, investigation of charge carrier dynamics with an ultrashort time scale is one of the primary steps necessary for developing high-speed devices. In the present study, we visualize ultrafast carrier dynamics in nanoscale devices using a combination of scanning photocurrent microscopy and time-resolved pump-probe techniques. We investigate transit times of carriers that are generated near one metallic electrode and subsequently transported toward the opposite electrode based on drift and diffusion motions. Carrier dynamics have been measured for various working conditions. In particular, the carrier velocities extracted from transit times increase for a larger negative gate bias, because of the increased field strength at the Schottky barrier.

*Electronic mail: ahny@ajou.ac.kr

The transit time of the charge carriers is a crucial factor limiting the high frequency response of nanoscale devices; however, traditional radio-frequency measurements are often limited by the high impedance or the RC constants of the devices[4-8]. Alternatively, optical ultrafast measurement techniques have been widely used to investigate charge carrier dynamics with a time resolution determined by the optical pulse width (down to a few femtoseconds)[9]. Recently, researchers have reported visualizing charge carrier movements in free-standing Si NWs using an ultrafast pump-probe imaging technique[10,11]. The carrier diffusion motions induced by a pump pulse located in the middle of the NWs were visualized; however, these optical measurements are limited for interrogating the carrier dynamics in operating devices because they strongly depend on the non-linear properties of materials and they are frequently obscured by the substrate signals. Consequently, these techniques are not ideal for low-dimensional systems with NWs thinner than the optical spot size (<100 nm) or with SWNTs.

Scanning photocurrent microscopy (SPCM) techniques have been introduced as powerful tools for investigating local optoelectronic characteristics, such as metallic contacts, defects, interfaces, and junctions[12-19]. We were able to collect localized electronic band information that is not disturbed by signals originating from the substrate, and hence, compared with conventional optical pump-probe techniques, SPCM can provide a higher signal-to-noise ratio. Only recently, ultrafast pump-probe photocurrent techniques have been demonstrated for studying carrier dynamics in carbon nanotube devices by using a collinear pump and probe beams, focused at the same position[20,21]. In addition, ultrafast phenomena in graphene and GaAs NWs have been investigated by measuring terahertz radiation that results from the ultrafast photocurrent in spatially designed circuit structures[22,23]. However, the carrier dynamics have not been directly visualized, in particular, with respect to various working conditions.

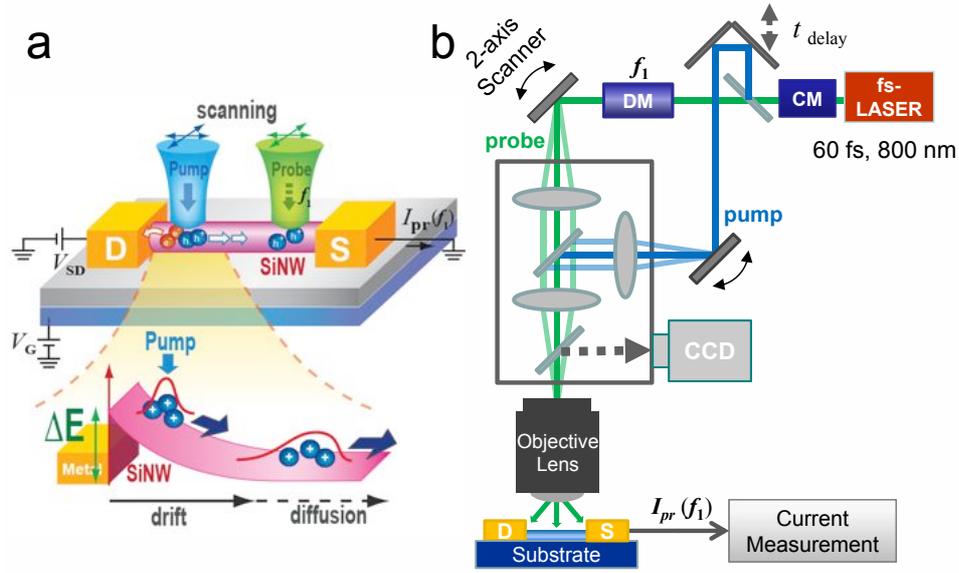

**Figure 1. Schematic of Experiments** (a) Schematic diagram of ultrafast carrier dynamics in a NW device. Photo-induced ultrafast carriers are generated by a femtosecond laser and are then transported toward the other part of the NWs by drift and subsequent diffusion processes. (b) Schematic diagram of the experimental setup. (CM: chirped mirrors, DM: deformable mirror)

Here we demonstrate a novel technique of femtosecond scanning photocurrent microscopy (fs-SPCM) on Si NWs and SWNT field-effect transistors (FETs). As schematically presented in Fig. 1a, ultrafast carrier dynamics in NW devices have been measured using a combination of scanning photocurrent microscopy[18,24] and time-resolved pump-probe techniques[9]. If we locate the femtosecond pump pulse near the metallic contacts (i.e., the Schottky contacts with strong electronic band bending), ultrashort carrier pulse will be generated and subsequently transported toward the middle of the NW by the drift motion resulting from built-in electric fields. Pure diffusion motions will follow once the carriers escape the depletion region. The carriers transported to the other part of the NWs can be monitored using the spatially separated probe pulse because the carriers subjected to pump-induced migration will influence the photocurrent generated by the probe pulse ($I_{pr}$). The value of $I_{pr}$ is recorded as a function of the position of both the pump and the probe pulses and also as a function of the time-delay between them, enabling us to obtain a spatio-temporal image of carrier movements. The experimental setup is schematically shown in Fig. 1b and its detailed description is presented in the method section.

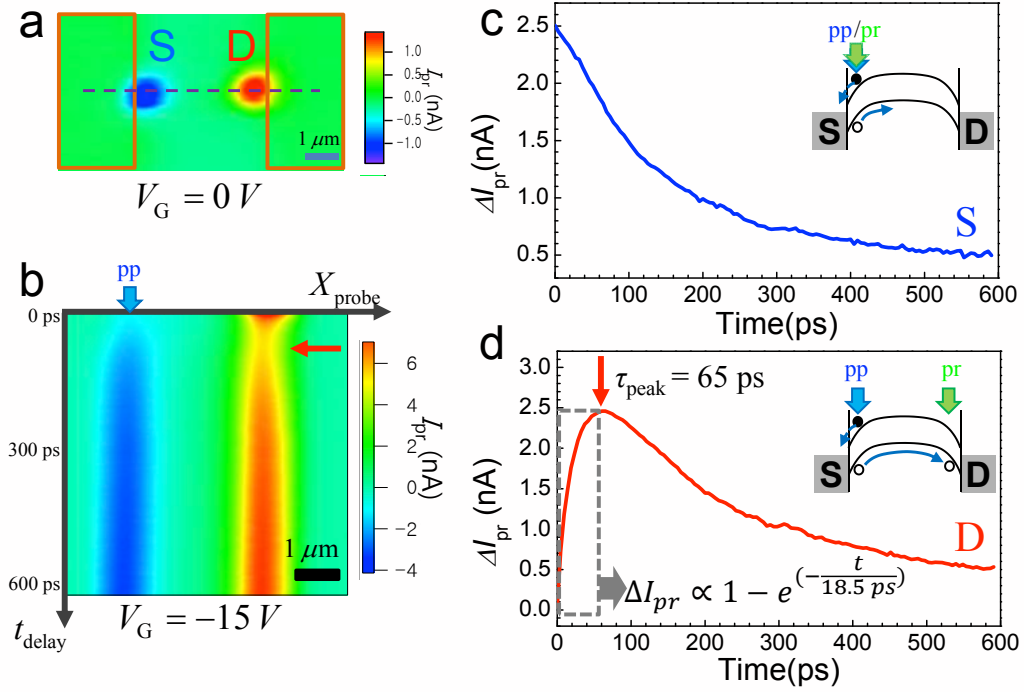

**Figure 2. Spatio-temporal Imaging of Photo-generated carriers** (a) A typical photocurrent microscopy image of a Si NW device with $l_{ch} = 4\,\mu$m without a pump pulse for $V_{SD} = V_G = 0$. (b) Plot of the spatio-temporal image obtained when the pump pulse is fixed at position S for $V_G = -15$ V. The x-axis represents the position along the dashed line in Figure 2a, and the y-axis represents time delay. The maximum change in signal $I_{pr}$ at position D is denoted by a red arrow. (c) Time trace of $\Delta I_{pr}$ when the probe is positioned at position S (when the probe spatially overlaps the pump pulse). (d) Time trace of $\Delta I_{pr}$ when the probe is positioned at D.

Figure 2a shows a typical photocurrent microscopic image of a Si NW device that has a channel length ($l_{ch}$) of $4\,\mu$m; this image is acquired by scanning the probe pulse (~1 kW/cm²) in the absence of the pump pulse, with the voltages for both the source-drain bias ($V_{SD}$) and the gate bias ($V_G$) fixed at 0 V. Photocurrent spots near the metal contacts are clearly visible along the Si NWs. Both spots are located approximately $0.25\ \mu$m from their respective metal interfaces, and this configuration results in a relative distance of $d_{pc} \sim 3.5\ \mu$m. In this figure, red (blue) indicates a positive (negative) current. The current spots near the metal electrodes originate from the electronic band bending at the electrode-NW interfaces, as reported in other studies[12]. Here, we are interested in visualizing the carrier movement created near one of the metallic contacts (denoted by S in Fig. 2a) and the subsequent migration to the other side of the electrode (denoted by D), which is positioned $3.5\ \mu$m from the pump position.

In the presence of a pump pulse (~10 kW/cm$^2$), $I_{pr}$ changes dramatically as a function of time delay ($t_{delay}$) for both the S and D signals. A series of $I_{pr}$ images as a function of $t_{delay}$ is provided in the Supplementary Information (S2). Here we demonstrate a two-dimensional plot of spatio-temporal imaging for a pump pulse at a fixed position (S) in Fig. 2b. Specifically, we measure $I_{pr}$ along the dashed line ($X_{probe}$) in Fig. 2a while varying $t_{delay}$ (0 − 600 ps). Both $I_{pr}$ signals (at positions S and D) decrease noticeably in the presence of the pump pulse. In particular, $I_{pr}$ is reduced significantly near the zero delay, which indicates that the change in $I_{pr}$ primarily results from the presence of the pump pulse.

More importantly, when the probe pulse is positioned at D, the maximum change in $I_{pr}$ is delayed, as indicated by a red arrow. The delayed change in $I_{pr}$ at position D is caused by the migrated carriers (generated by the pump pulse), which are spatially separated at position S. When the electron and hole pairs are created in the band-bending region, the electrons move to the nearby metal electrode (S), while the holes are injected toward the opposite electrode (D) in p-type operation, as shown in the insets of Fig. 2c and 2d. The migrated hole carriers occupy some of the states that would have been excited by the probe pulse, and this phenomenon results in the decreased $I_{pr}$.

The magnitude of the change in $I_{pr}$ ($\Delta I_{pr}$) for positions S and D is plotted as a function of $t_{delay}$ in Figs. 2c and 2d, respectively. For position S, $\Delta I_{pr}$ is largest at zero delay and gradually relaxes with a decay constant of 140 ps. The measurement of the carrier escape time for photogenerated carriers has been reported for SWNT devices by using the collinear pump-probe photocurrent technique[21]. Conversely, when we measure $\Delta I_{pr}$ on the opposite side of the electrode (at position D), $\Delta I_{pr}$ increases rapidly until it reaches its highest value at $\tau \sim 65\,\text{ps}$. This transit time ($\tau_{peak}$) corresponds to a carrier velocity of $5.4 \times 10^6$ cm/s. In addition, the initial increase in $\Delta I_{pr}$ for $t_{delay} < \tau_{peak}$ is well fitted by the relation $\Delta I_{pr} \propto 1 - exp(-t_{delay}/\tau_c)$, where $\tau_c$ is the characteristic time constant (when $\Delta I_{pr}$ reaches 63% of its maximum value), which yields a value of 18.5 ps. Similar results on the different channel lengths are presented in the Supplementary Information (S3). We also note that carrier dynamics can be obtained for the probe pulse located in the middle of NWs, even though this visualization is not provided in Fig. 2 because of the low contrast ratio relative to the metallic signals (Supplementary Information S4).

As previously mentioned, fs-SPCM is a powerful technique for investigating carrier dynamics for nanoscale systems with a very low cross-section, such as thin NWs and SWNTs. The results for the Si NWs with a diameter of 30 nm ($\tau_{peak} \sim 16$ ps for 1-$\mu$m separation) and for the SWNT ($\tau_{peak} \sim 5$ ps for 4-$\mu$m separation) are shown in the Supplementary Information (S6 and S7). In particular, the transit time of SWNT device is about 10 times shorter than that of Si NWs with similar channel length, because of the higher SWNT mobility. We also introduce a technique that measure the probe photocurrent as a function of both the pump and the probe positions as discussed in Supplementary Information (S5). This is particularly useful for identifying pump locations where nanoscale devices experience interesting dynamical phenomena.

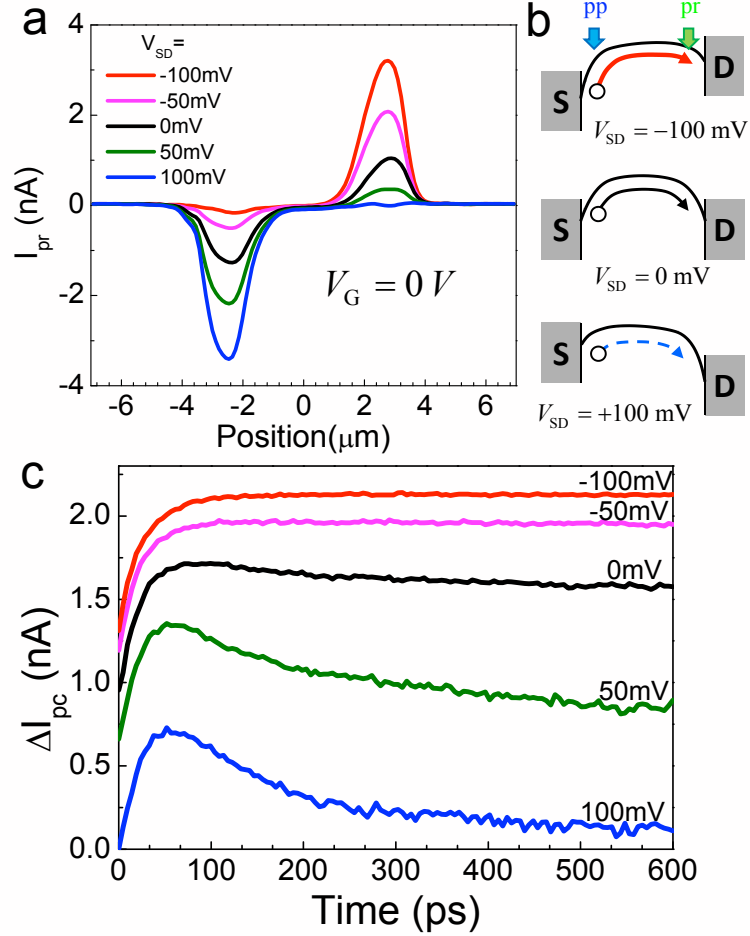

**Figure 3. Source-drain Bias Voltage Dependence** (a) Line profiles of $I_{pr}$ along the Si NW axis without pump pulse for a device with $l_{ch} = 6.5\ \mu m$, for the five different $V_{SD}$'s. PC spots are localized near the metallic contacts regardless of $V_{SD}$ for low bias conditions, whereas the strength of S and D signals varies with $V_{SD}$ (b) Schematic representation of the electrostatic potentials extracted from (a). (c) Plot of $\Delta I_{pr}$ as a function of time delay with the presence of the pump pulse at the position S for the five different $V_{SD}$'s.

It is crucial to address dynamical carrier movements in terms of various working conditions, such as the source-drain bias and the gate bias, which are configurations that strongly modify the electronic band structures of NWs. We first address carrier dynamics with respect to source-drain bias voltage ($V_{SD}$) for a device with $l_{ch} = 6\,\mu m$. We show in Fig. 3a a line profile of $I_{pr}$ without the pump pulse, for the five different $V_{SD}$'s from −100 mV to 100 mV. For the relatively low bias conditions, the potential drop is dominant near the metal contacts, as schematically shown in Fig. 3b. We measured $I_{pr}$ at the opposite electrode D as a function of $t_{delay}$ for the fixed pump position at S. Figure 3c shows $\Delta I_{pr}$ as a function of $t_{delay}$. We found that $\Delta I_{pr}$ decreases rapidly for $V_{SD} > 0$, whereas $\Delta I_{pr}$ persists for $V_{SD} < 0$. This is because, for $V_{SD} < 0$, the electrostatic potential is lower at the probe position (D) than that of the pump position (S) with respect to hole carriers. Conversely, for $V_{SD} > 0$, the potential at the opposite electrode is higher and this reduces the number of the carriers that arrive at the opposite electrode. In other words, only the carriers with higher kinetic energies can reach the opposite electrode, resulting in the narrowing of the electrical pulse width.

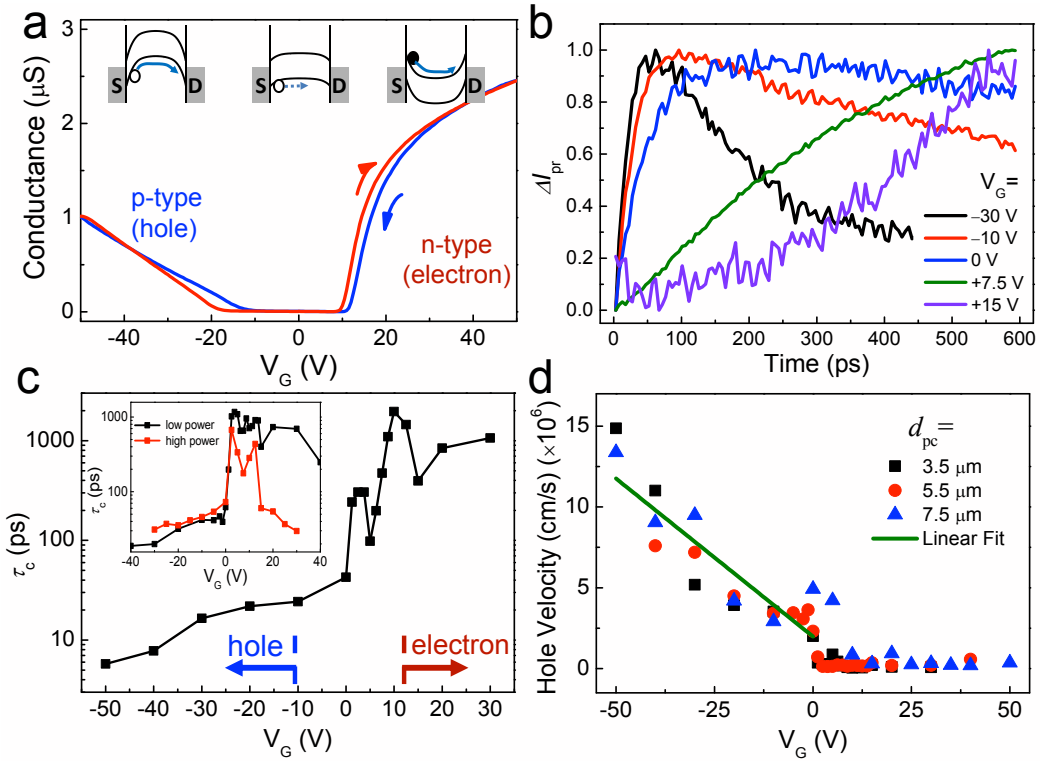

**Figure 4. Gate-dependent Carrier Transport Phenomena** (a) DC conductance as a function of $V_G$ for the sample with $d_{pc} = 3.5\ \mu\text{m}$ with a relatively low pump intensity of $I_0 = 10\ \text{kW/cm}^2$. High on-state current and the large transconductance for n-type region result from Si electron mobility, which is higher than the hole mobility. Band-gap region ranges from −12.2 to +11.5 V; gate efficiency parameter $\alpha$ is 0.047 (Insets) Schematic diagrams of the gate-dependent band alignments. Electric field strength in depletion region decreases as $V_G$ increases in p-type operation (from $-50\ \text{V}$ to $0\ \text{V}$); consequently initial speed of the injected carriers decreases accordingly. (b) Normalized time traces of $\Delta I_{pr}$ for various values of $V_G$. (c) $\tau_c$ as a function of $V_G$. $\tau_c$ increase from $6\ \text{ps}$ ($V_G\ 50\ \text{V}$) to $43\ \text{ps}$ ($V_G = 0\ \text{V}$) (Inset) $\tau_c$ as a function of $V_G$ with $d_{pc} = 5.5\ \mu\text{m}$ for both low- ($I_0$) and high-intensity ($10\ I_0$) conditions. (d) $v_{peak}$ as a function of $V_G$ for the three devices with different values of $d_{pc}$ (3.5, 5.5, and 7.5 $\mu$m). The solid line is a linear fit for the data with $V_G < 0$.

Figure 4a shows the DC conductivity of the device ($l_{ch} = 4$ μm) as a function of $V_G$. Fig. 4b shows a plot of $\Delta I_{pr}$ as a function of $t_{delay}$ for the four different gate-bias configurations. The transit time increases significantly as we increase $V_G$ from −30 V to 15 V. The time constants $\tau_c$ were obtained by fitting the time traces and the results are plotted in Fig. 4c. In particular, $\tau_c$ increases rapidly for $V_G > 0$ V, and this increase reflects the significant reduction in the band-bending strength in the NWs (i.e., the suppressed drift motions). It is likely that carrier transport is dominated by a hole diffusion process for $V_G > 10$ V.

Because the band-bending direction is reversed for $V_G > 10\text{V}$, the electron carriers (instead of the hole carriers) are injected toward the opposite electrode. Therefore, it is surprising that we cannot observe experimental evidence regarding electron transport phenomena in the n-type operating condition. This limitation likely occurs because we use a device with moderately n-doped materials; specifically, our technique is best suited for interrogating the dynamics of the minority carrier[25]. However, if we increase the pump intensity by a factor of ten (as shown in the inset of Fig. 4c), a dramatic reduction in $\tau_c$ is observed for n-type operation (e.g., $\tau_c \sim 30$ ps at $V_G = 30$ V for the sample with $d_{pc} \sim 5.5$ $\mu$m), and this finding likely results from the transport of electrons.

Corresponding to the maximum $\Delta I_{pr}$, the most probable hole velocity ($v_{peak}$) in Fig. 4b is plotted as a function of $V_G$, as shown in Fig. 4d. For this plot, we use the gate-dependent transit times obtained from Fig. 4c and the following relation $\tau_{peak} \approx 3.5 \tau_c$. We combine the results for three devices with different channel lengths of 4, 6, and 8 $\mu$m (with $d_{pc}$ of 3.5, 5.5, and 7.5 $\mu$m, respectively). The velocity decreases linearly as the field strength is reduced, and this result is in accordance with the change in the gate potential (from $V_G = -50$ V to $V_G = 0$ V). This finding is expected because of the field-dependent drift velocity of $\vec{v} = \mu_h \vec{E}$, where $\mu_h$ is the hole mobility and $\vec{E}$ represents the localized electric fields. By fitting the data and applying the gate efficiency parameter, we extracted a minority hole mobility of 200 cm$^2$/Vs, which is in reasonable agreement with the known value[26,27].

It is interesting that the gate-dependent velocity measured at opposite electrodes does not change noticeably for different channel lengths. This finding is a significant deviation from the ordinary diffusion model, as derived from simulation results with a transient drift-diffusion model, which predicts a drastic decrease in $v_{peak}$ when $l_{ch}$ is larger than the width of the depletion layer (500 nm) (Supplementary Information S8). In general, the carrier velocity tends to decrease rapidly when carriers are subject to a pure diffusion process outside the depletion region. However, our experimental data suggest that the average carrier velocity is independent of $l_{ch}$. This finding is described as drift-like motion, which has been reported previously in time-of-flight measurements[28]. Attributed to both surface recombination effects and the unique transport properties at high carrier kinetic energies, this phenomenon necessitates future study using a kinetic model that includes non-equilibrium initial carrier distributions.

In summary, we developed a novel technique based on ultrafast photocurrent microscopy, by combining scanning photocurrent microscopy and ultrafast pump-probe techniques. The transport of carriers created by femtoseconds pump pulse has been imaged by observing the change in the probe photocurrent. In particular, we measured the transit time of the carriers transported to opposite electrodes. We studied carrier dynamics in various working conditions, such as source-drain and gate biases. Gate-dependent measurements reveal that the carrier velocity changes linearly with the applied gate bias in accordance with changes in the electric field strength in the Schottky barrier. We observed drift-like motion, in which the average velocity did not change noticeably with changes in the channel length (up to $8\,\mu\text{m}$). This work represents an important step toward understanding ultrafast dynamics in various nanoscale devices and toward developing future high-speed electronic devices.

**Methods**

**Femtosecond Scanning Photocurrent Microscopy:** A femtosecond Ti:Sapphire laser (centered at 800 nm with a repetition rate of 80 MHz and a pulse width of 60 fs) is used to locally photoexcite a NW device. The laser beams were divided into pump and probe beams, and $t_{delay}$ is generated by a delay stage, which is located in the path of the pump beam. A pair of chirped mirrors (CM) is used to compensate for the positive dispersion that originates mainly from the objective lens. The pump and probe pulses are focused on the samples using the objective lens (100X and N.A. 0.80), which has a 600-nm spatial resolution. A pair of two-axis steering mirrors (Newport Corporation, Inc.) is used to manipulate the positions of both focused laser spots. An optical modulator (Boston Micromachines Corporation) is used to optically modulate the probe pulse at 20 kHz to capture only the $I_{pr}$ signals and to exclude the photocurrents generated directly by the pump pulse. $I_{pr}$ is measured using a high-speed current preamplifier (FEMTO Messtechnik GmbH) and a subsequent lock-in amplifier (Signal Recovery) operated at the probe modulation frequency.

**Device Fabrication:** Si NW FETs shown in the main text are fabricated from silicon-on-insulator wafers (Soitec, Inc.), which have an oxide layer (1 $\mu$m) and a moderately doped n-type Si layer (thickness of 200 nm and doping concentration of $1.6 \times 10^{16}$ /cm$^3$). NW patterns are generated with widths of 200 − 300 nm and channel lengths of 4 –10 μm, by an electron-beam lithography technique followed by inductively coupled plasma (ICP) etching.[29,30] Etched Si NWs are then electrically connected by Ni and Au (50 nm and 200 nm, respectively) electrodes using electron-beam lithography. After these processes, the device is annealed at 250°C for 10 min under an Ar atmosphere.

**Acknowledgements** This work was supported by Mid-career Researcher Programs (2011-0016173) and by PRC Program (2009-0094046) through National Research Foundation grant funded by the Korea Government.

**Author Contributions** Y.H.A conceived and designed the experiments. B.H.S and J.K.P fabricated devices and performed experiments. J.T.H participated in data analysis. All authors including J.Y.P and S.L. discussed the results and participated in writing the manuscript.

**Competing Interests Statement** The authors declare that they have no competing financial interests.

**Correspondence.** Correspondence and requests for materials should be addressed to Y.H.A (ahny@ajou.ac.kr)

**Supporting Information for**

Imaging ultrafast carrier transport in nanoscale devices using femtosecond photocurrent microscopy

*B. H. Son, J. K. Park, J. T. Hong, J. Y. Park, S. Lee, and Y. H. Ahn**

*Department of Physics and Department of Energy Systems Research, Ajou University, Suwon 443-749, Korea*

### S1. SEM image of a Sample

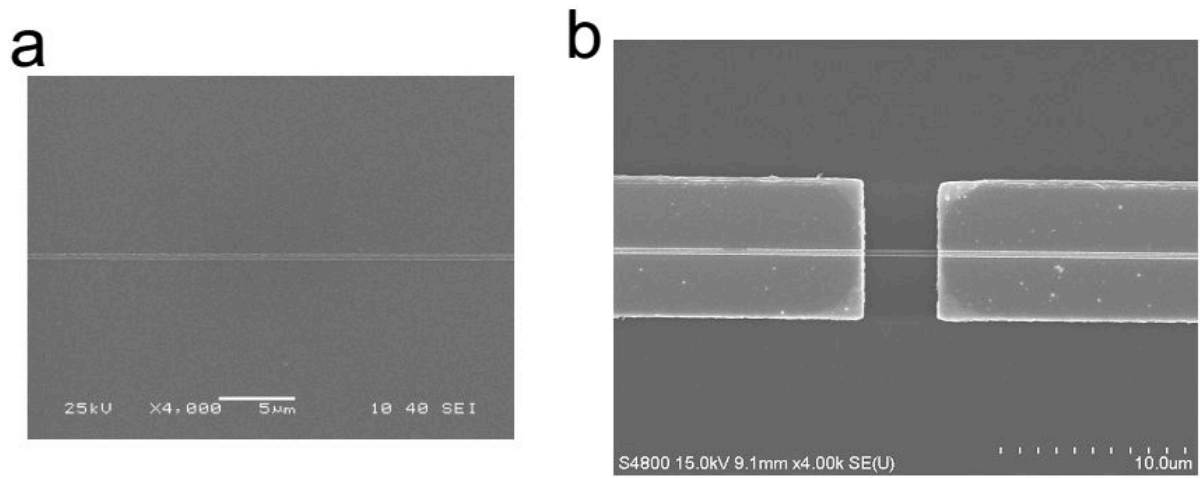

**Fig. S1** (a) SEM image of an isotropically etched silicon nanowire by reactive ion etching. (b) SEM image of a Si NW device with Ni/Au (50/200 nm, respectively). The scale bars are 5 μm and 10 μm, respectively.

Our devices were fabricated on a silicon-on-insulator wafer (Soitec, Inc.) which consists of 1 μm $SiO_2$ layer and 200 nm moderately-doped, n-type silicon layer with a doping concentration of $1.6 \times 10^{16}$ /cm$^3$. Nanowire pattern that has a width of 200 nm was generated by electron beam lithography techniques. A negative tone photoresist, ma-N 2400 series, was used to form etching mask. The reactive ion etching process with oxygen and tetrafluoromethane was followed to define top-down fabricated silicon nanowires (Si NW) as shown in an SEM image of Fig. S1a. The Si NWs were then electrically connected by Ni/Au (50 and 200 nm, respectively) electrodes by an e-beam lithography with PMMA as a positive e-beam resist. The distance between the source and drain electrodes was in range of 4 –10 μm. After these processes, the device was annealed at 250 ˚C for 10 min under Ar atmosphere to improve contact properties between Si NW and the metal film.

## S2. 2D probe-photocurrent ($I_{pr}$) image as a function of time delay

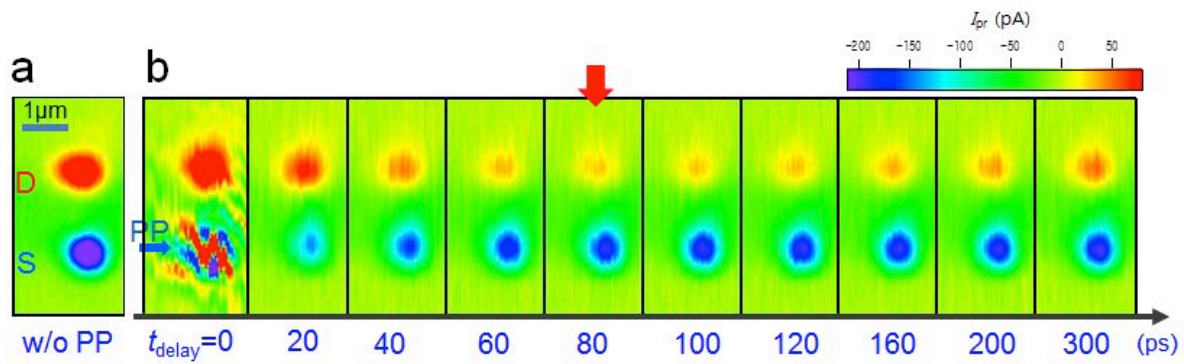

**Fig. S2** (a) Probe-photocurrent $I_{pr}$ image without pump pulse (b) A series of 2D probe-photocurrent images as a function of the time delay, when pump pulse is located at S position for a sample with $l_{ch} = 4\ \mu m$.

Fig. S2a shows a probe-photocurrent $I_{pr}$ image without pump pulse for the device shown in Fig. 2 of the main manuscript ($l_{ch} = 4\ \mu$m). Fig. S2b is a series of $I_{pr}$ images as a function of time-delay ($t_{delay}$) for the pump pulse located at S position. With the presence of pump pulse (~30 kW/cm$^2$), $I_{pr}$ changes dramatically for both S (blue) and D (red) signals, exhibiting clear dynamic behaviors. Both $I_{pr}$ signals (at S and D positions) decrease noticeably with the presence of the pump pulse. In particular, $I_{pr}$ is reduced significantly near the zero delay, whereas the maximum change in $I_{pr}$ is delayed when the probe pulse is located at position D as depicted by a red arrow.

## S3. Spatio-temporal image for a sample with a different channel length

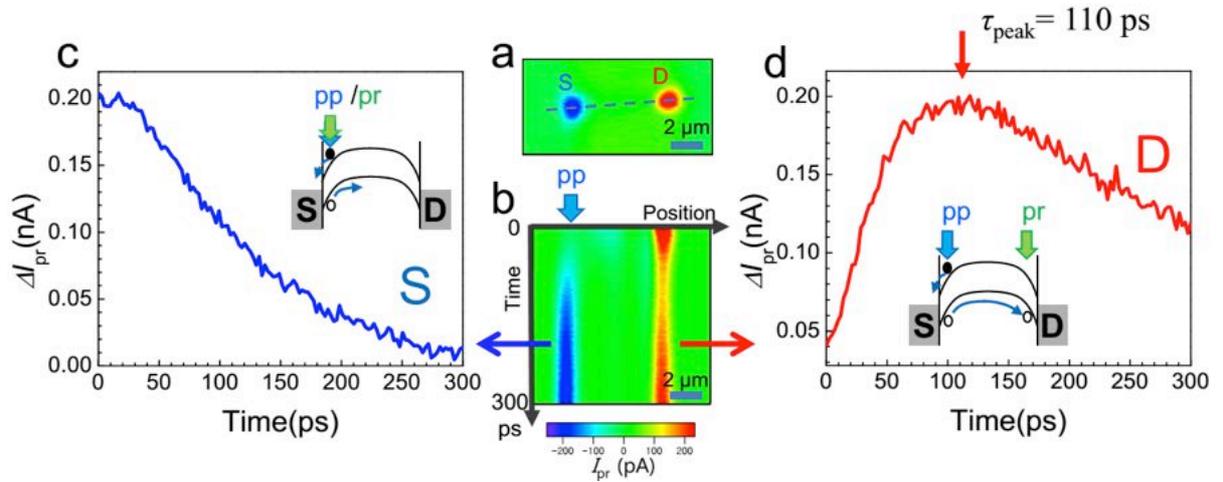

Fig. S3 (a) $I_{pr}$ for the samples with $l_{ch} = 6.5\ \mu m$, without pump pulse (b) Spatio-temporal image along the dashed line in (a) when the pump pulse is positioned at S. (c) $\Delta I_{pr}$ as a function of time delay when the pump and the probe pulse have spatial overlap. (d) That of (c) with the probe pulse positioned at the opposite electrode.

We recorded the spatio-temporal images for the devices with a relatively long channel length of $l_{ch} = 6.5\ \mu m$ (with $d_{sp}$ : 6.0 $\mu m$) for $V_{SD} = 0$ V and $V_G = -10$ V. As in the case of the device shown in Fig. 2 of the main manuscript, the change in the probe-photocurrent ($\Delta I_{pr}$) is largest at zero delay for position S, whereas when we measure $\Delta I_{pr}$ on the opposite side of the electrode (at position D), $\Delta I_{pr}$ increase rapidly until it reaches the highest value at $\tau_{peak} \sim 110$ ps. We note however that this transit time depends strongly on the gate bias voltage as discussed in Fig. 4 of the main manuscript.

## S4. Pump-induced Photocurrent Changes in the Middle of NWs

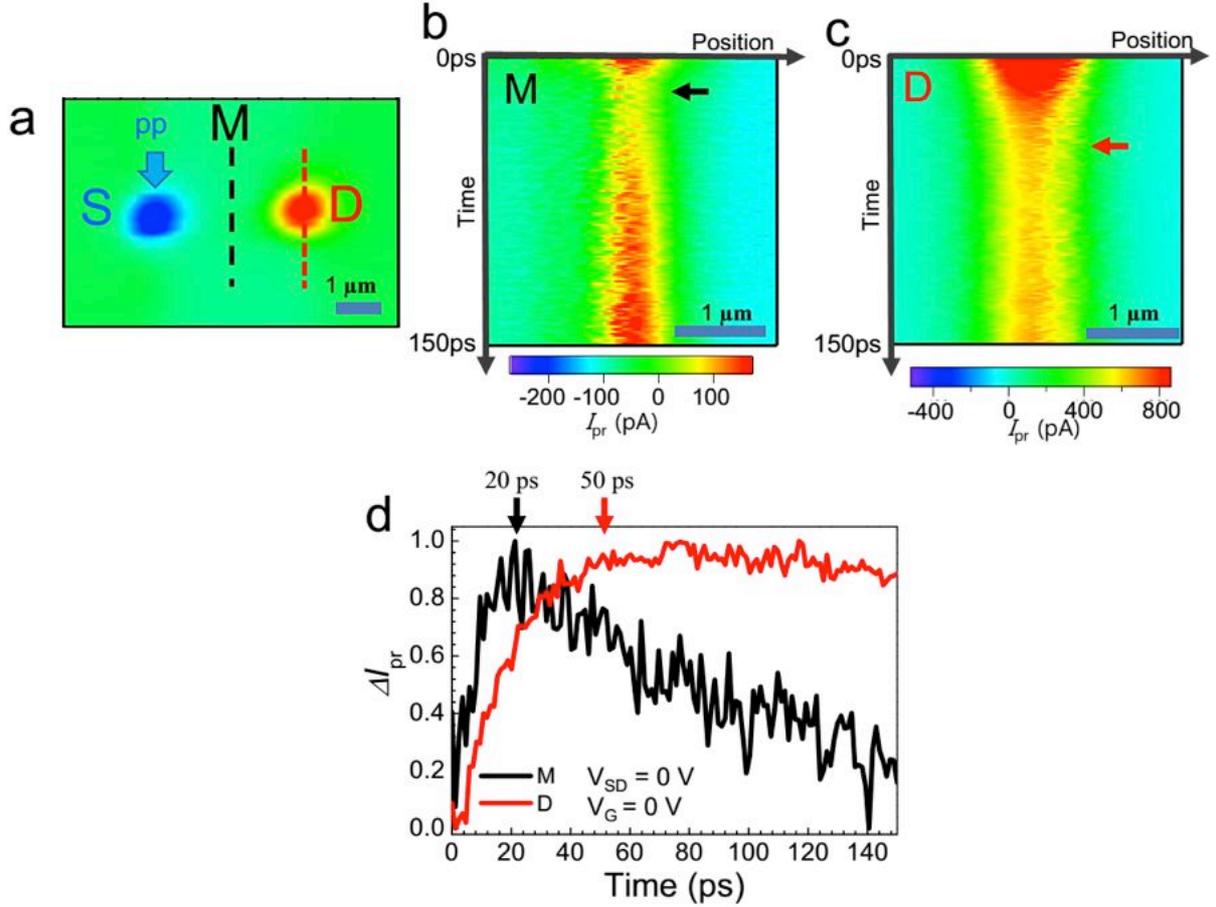

Fig. S4. (a) An $I_{pr}$ image without the pump pulse for a Si NW device with $l_{ch} = 4\ \mu m$. (b) A spatio-temporal images when the probe pulse is scanned along the dashed line (M). (c) That of (b) when the probe pulse is scanned along the dotted line (D). (d) Plot of $\Delta I_{pr}$ as a function of time obtained from (b) (black lien) and (c) (red line).

Because $I_{pr}$ in the middle of NWs is generally overwhelmed by the metallic signals, the carrier movements are not clearly resolved in a spatio-temporal image shown in Fig 2 of the main manuscript. Here, we scan the probe pulse in a direction perpendicular to the NW axis, e.g., along a dashed line in the pump-free $I_{pr}$ image of Fig. S4a. Fig. S4b and S4c show spatio-temporal images when the probe pulse is scanned along the dashed (M) and dotted lines (D) in Fig. S4a, respectively. As summarized in the plots of Fig. S4c, the transit time (~20 ps) measured at the center of NWs which is shorter than that of the opposite electrodes (~50 ps). In other words, the transit times can be measured along NWs as long as the probe pulse does not have a noticeable spatial overlap with the metal contact region.

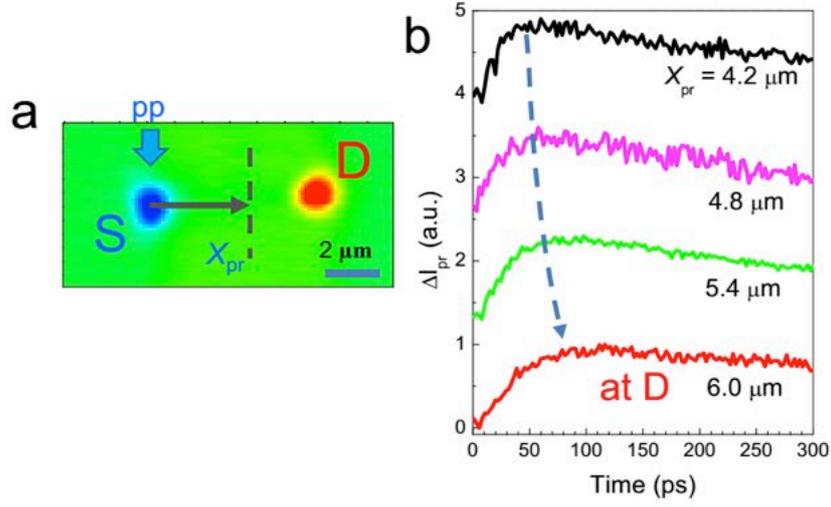

Fig. S5. (a) $I_{pr}$ image without the pump pulse for a Si NW device with $l_{ch} = 6.5\ \mu m$. (b) Plot of $\Delta I_{pr}$ as a function of time for various distance $X_{pr}$ from the pump pulse location (position S).

Another example is shown in Fig. S5 for the device with a longer channel length ($l_{ch} = 6.5\ \mu m$). Fig. S5a shows a pump-free $I_{pr}$ image and Fig. S5b shows $\Delta I_{pr}$ as a function of $t_{delay}$ for various distance $X_{pr}$ from the pump pulse location (position S). They show a gradual increase in $\tau_{peak}$ with $X_{pr}$.

## S5. Pump-probe Correlation Imaging

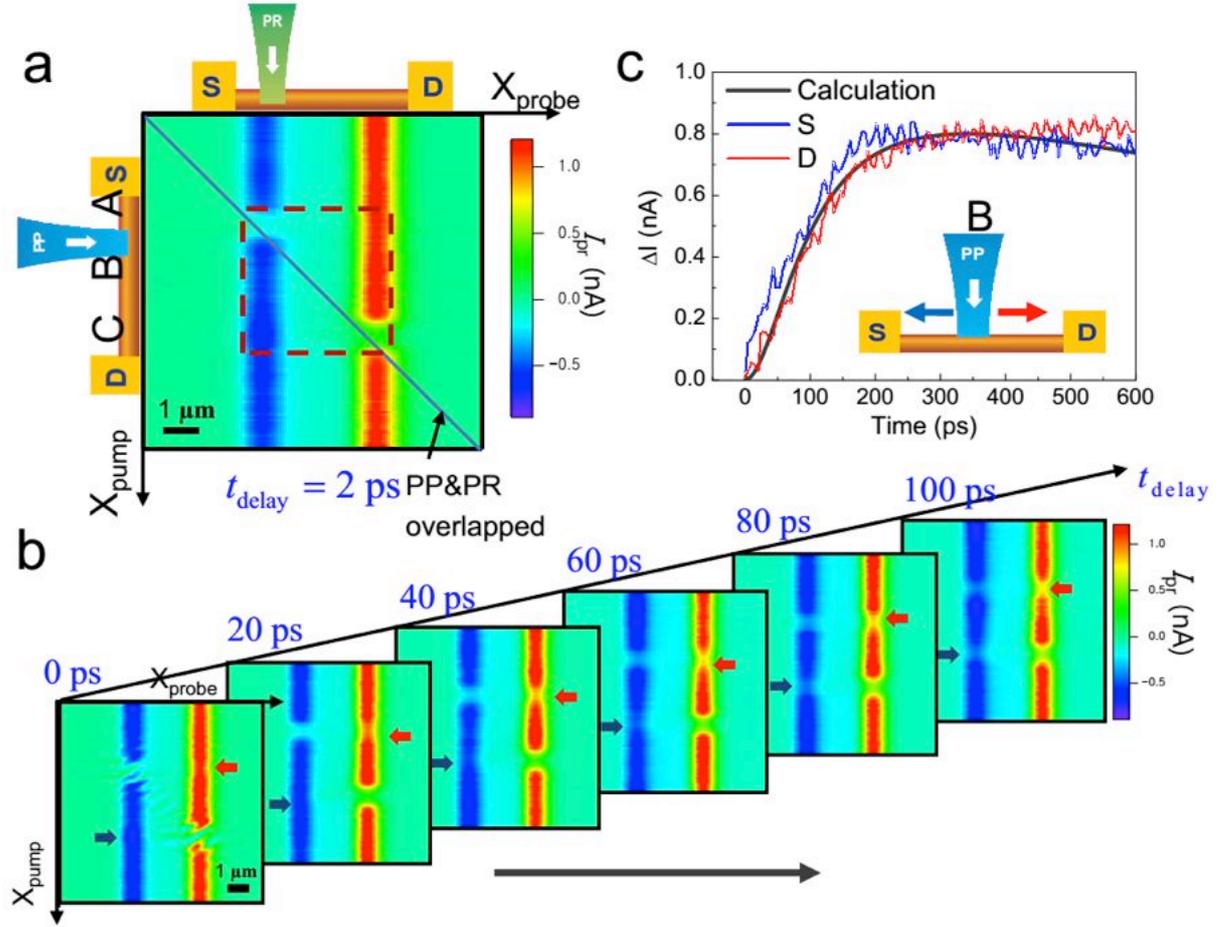

**Fig. S6.** (a) Probe photocurrents are mapped as function of probe ($X_{probe}$) and pump ($X_{pump}$) positions for $V_G$ = 0 V and $t_{delay}$ = 2 ps. (b) Three-dimensional spatio-temporal image, which is constructed by accumulating the two-dimensional images shown in (a) as a function of $t_{delay}$ (from 0 ps to 100 ps). (c) Plots of $\Delta I_{pr}$ measured at positions S (blue line) and D (red line) as a function of $t_{delay}$ when the pump pulse illuminates the center of the NW, as indicated by position B in (a). The black solid line indicates the results of the numerical calculation based on a transient drift-diffusion model.

Using the techniques presented in this paper, carrier dynamics based on a fixed pump position (located at the metallic contact region) have been successfully demonstrated by recording the value of $I_{pr}$ as a function of the probe position. Here, we address carrier dynamics as a function of both the pump and the probe positions by introducing a novel technique based on pump-probe photocurrent correlation images. A representative image measured at $V_G = 0$ V, and $t_{delay} = 2$ ps is shown in Fig. S6a. $X_{probe}$ and $X_{pump}$ represent the positions of the probe and the pump spots, respectively, and therefore, the solid diagonal line indicates the region where the two pulses spatially overlap. This image is particularly useful for identifying pump locations where nanoscale devices experience interesting dynamical phenomena. A dramatic decrease in $I_{pr}$ is clearly visible where there is a spatial overlap between the pump and the probe pulses at approximately zero delay. As expected, $\Delta I_{pr}$ is noticeable when the pump pulse is positioned near the metallic contacts (denoted by A and C).

A three-dimensional spatio-temporal image can be constructed by accumulating two-dimensional pump-probe correlation images as a function of $t_{delay}$. A series of correlation images is shown in Fig. S6b as a function of $t_{delay}$, ranging from 0 ps to 100 ps, with a 20-ps interval. When the pump pulse illuminates a band-bending region around A or C, the dynamical behaviors of $I_{pr}$ are consistent with the behaviors in Fig. 2 of the main text. In particular, we observe an abrupt decrease in $I_{pr}$ at $t_{delay} = 0$ ps when the pump and probe pulses are spatially overlapped (diagonal positions) and a delayed change in $I_{pr}$ at opposite electrodes, as indicated by red and blue arrows (off-diagonal positions).

Interestingly, we can also observe the dynamical change in $I_{pr}$ when the pump pulse illuminates a flat-band region [i.e., the middle of the NW (denoted by B in Fig. S6a)]. This phenomenon occurs when pump-induced carrier dynamics are dominated by the pure diffusion process because there is no built-in electric field. In Fig. S6c, we show a plot of $\Delta I_{pr}$ measured at the metallic contacts (positions S and D) as a function of $t_{delay}$ when $X_{pump}$ is located at the center of the NW. We found a gradual increase in $\Delta I_{pr}$ as a function of $t_{delay}$, which likely results from holes migrating through the diffusion process. The $\tau_{peak}$ value of ~200 ps is in good agreement with value generated by numerical calculation (represented by a black solid line) using the time-dependent carrier diffusion equation (Supplementary Information S8).

## S6. Transit time measurements on SWNT devices.

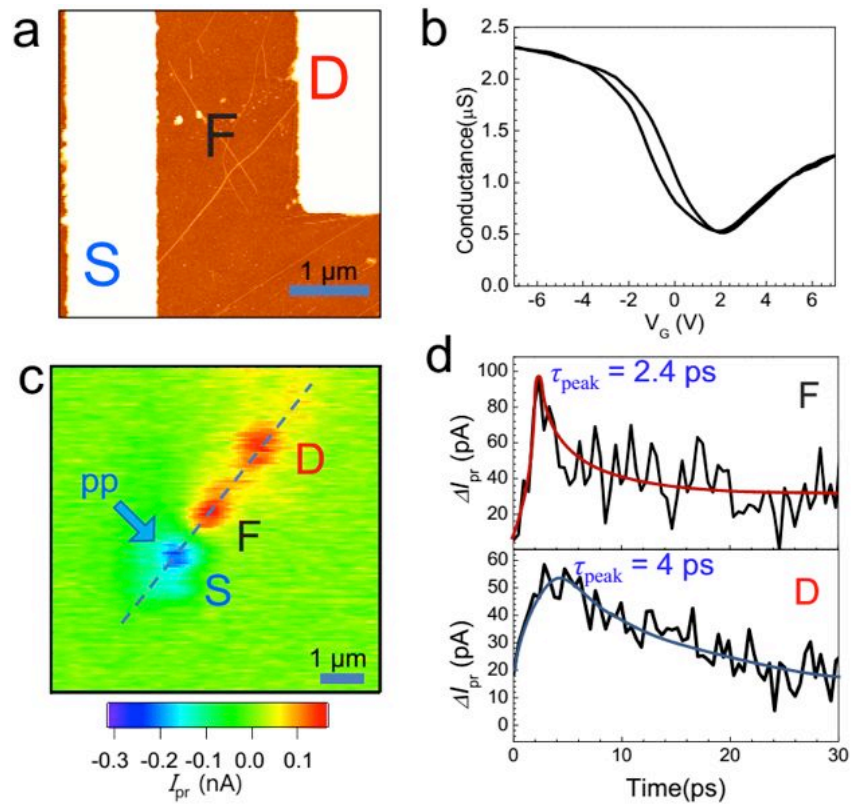

Fig. S7 (a) AFM image of a SWNT device. (b) DC conductance as a function of gate bias voltage. (c) SPCM image for $I_{pr}$ without the presence of pump pulse (d) $\Delta I_{pr}$ is plotted as a function of time delay for probe position at F (defect) and D (opposite electrode) spots, from top to bottom, with the fixed pump position at S spot. Red and blue lines are guidelines for clarity.

In the main manuscript, we demonstrated the top-down fabricated Si NW devices particularly because of their low hysteric behavior with respect to the gate response, which help us to obtain consistent results especially to address the gate-dependent carrier dynamics. However, our technique is still powerful for interrogating the devices with a very low optical cross-section, such as with thin NWs (supporting information S7) and with SWNTs.

We fabricated FET device on SWNT synthesized by chemical vapor deposition (CVD) methods as reported elsewhere[1,2]. Fig. S7a shows an AFM image of the device whose DC conductance is plotted as a function of gate bias voltage in Fig. S7b. The standard SPCM image of $I_{pr}$ (~45 kW/cm$^2$) without the presence of the pump pulse is shown in Fig. S7c. In addition to the metallic PC signals (denoted by S and D, separated by $d_{PC} \sim 3 \ \mu m$), it also shows a clear defect signal in the middle of SWNT (denoted by F), which is located ~1 $\mu m$ from the position S. With the fixed pump pulse (~170 kW/cm$^2$) at position S, we measured $I_{pr}$ as a function of time delay at the two different positions of the defect (B) and near the opposite electrode (D) as shown in Fig. S7d. The transit times of 2.4 ps and 4.0 ps have been obtained for F and D positions, respectively. The most probable velocity of carriers that reach the opposite electrode corresponds to $7.5 \times 10^7$ cm/s, which is more than 10 times faster than that of Si NWs shown in the main text.

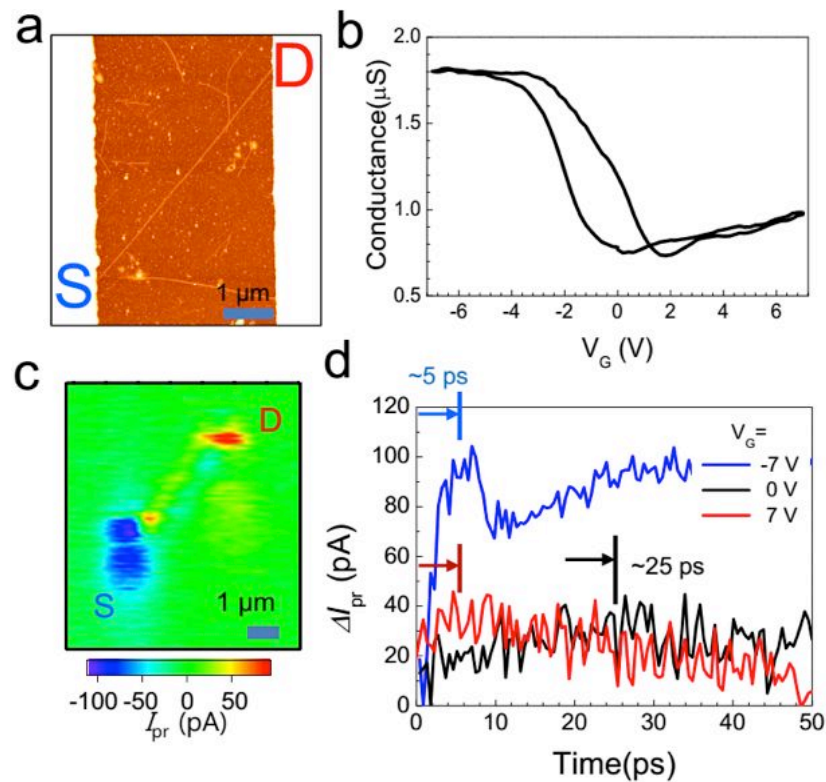

Fig. S8 (a) AFM image of a SWNT device. (b) DC conductance as a function of gate bias voltage. (c) SPCM image for $I_{pr}$ without the presence of pump pulse (d) $\Delta I_{pr}$ plotted as a function of time delay for the three different gate bias voltages of $V_G = -7$ V (blue line), 0 V (grey line), and 7 V (red line), with the fixed pump position at S position.

In Fig. S8, we show the gate-dependent transit times for another SWNT device with $d_{pc} \sim 5.5\ \mu m$. Fig. S8a shows an AFM image of the SWNT device whose DC conductance is plotted as a function of gate bias voltage in Fig. S8b. The standard SPCM image of $I_{pr}$ without the presence of the pump pulse is shown in Fig. S8c. $\Delta I_{pr}$ is plotted in Fig. S8d for the three different gate bias voltages of $V_G = -7$ V (blue line), 0 V (grey line), and 7 V (red line). The carrier velocities are apparently faster in p-type (hole) and n-type (electron) operation than that of $V_G = 0$ V. The higher velocity for the large gate bias voltage is likely due to the increased band-bending strengths as discussed in the main text.

## S7. Results on CVD-grown Si NWs

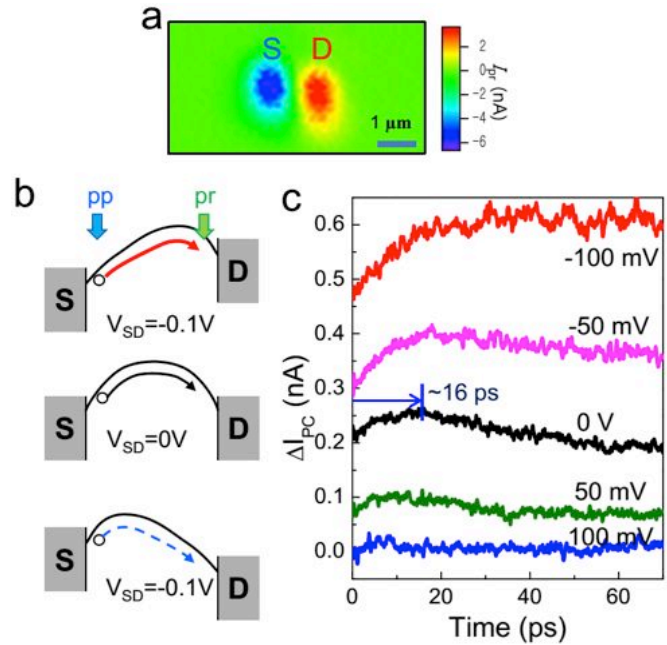

Fig. S9 (a) An $I_{pr}$ image without the pump pulse for a Si NW device with a diameter of 30 nm and with $l_{ch} = 2\ \mu m$. (b) Schematic representation of the electrostatic potentials for different $V_{SD}$'s. (c) $\Delta I_{pr}$ is plotted as a function of time for different $V_{SD}$'s, when the pump (probe) pulse is located at position S (D).

Thin Si NWs fabricated by a chemical vapor deposition method[3] demonstrate similar behaviors as in the case of the top-down fabricated devices shown in the main manuscript. We demonstrate here the results on p-type Si NWs with a diameter of 30 nm. The channel length is also very short ( $l_{ch} = 2\,\mu$m ) making the relative distance between S and D, $d_{sp} \sim 1\,\mu$m as shown in Fig. S9a. Therefore it is likely that there is no flat band region in the middle of NWs. For the fixed pump position at S, we measured $I_{pr}$ at the opposite electrode D as a function of time delay. We used relatively high powers of 400 kW/cm$^2$ and 30 kW/cm$^2$ for the pump and the probe pulses, respectively. Fig. S9b shows $\Delta I_{pr}$ as a function of time delay for the different $V_{SD}$'s from −100 mV to 100 mV, for the fixed gate bias of $V_G = 0$ V. At $V_{SD} = 0$ V (middle of Fig. S9a), the transit time ($\tau_{peak}$) is measured at ~ 16 ps, which is apparently shorter than that of the Si NW device with $d_{sp} = 3.5\,\mu$m, because of the short channel length. It is very clear that, $V_{SD} < 0$, $\Delta I_{pr}$ persists very long time, whereas decreases rapidly for $V_{SD} > 0$ as shown previously in Fig. S6. As mentioned before, this is because for $V_{SD} > 0$, the potential with respect to hole carriers is higher at the opposite electrode than that of the pump position, and this results in the blocking of carrier movements. In other words, hole carriers only with the higher kinetic energy can reach the opposite electrode efficiently.

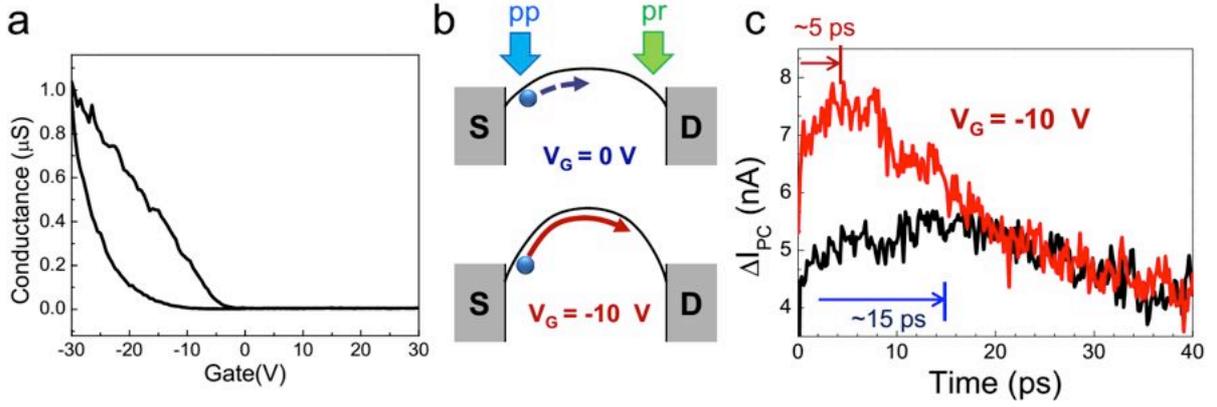

Fig. S10 (a) DC conductance as a function $V_G$ for a Si NW with a diameter of 30 nm and with $l_{ch} = 2\ \mu m$. (b) Schematic representation of the electrostatic potentials for different gate biases. (c) $\Delta I_{pr}$ is plotted as a function of time for $V_G = 0$ V (black line) and $-10$ V (red line) when the pump (probe) pulse is located at position S (D).

Although the device suffers from the hysteric effects as shown in the DC-conductance plot of Fig. S10a, we can also observe that the transit time increases for the large negative bias if measured sufficiently fast (less than 10 s). Fig. S10c shows $\Delta I_{pr}$ as a function of time delay for $V_G = 0$ V (black line) and $-10$ V (red line), with zero source-drain bias. The transit time was about three time shorter for $V_G = -10$ V than for $V_G = 0$ V.

## S8. Analysis based on transient drift-diffusion model

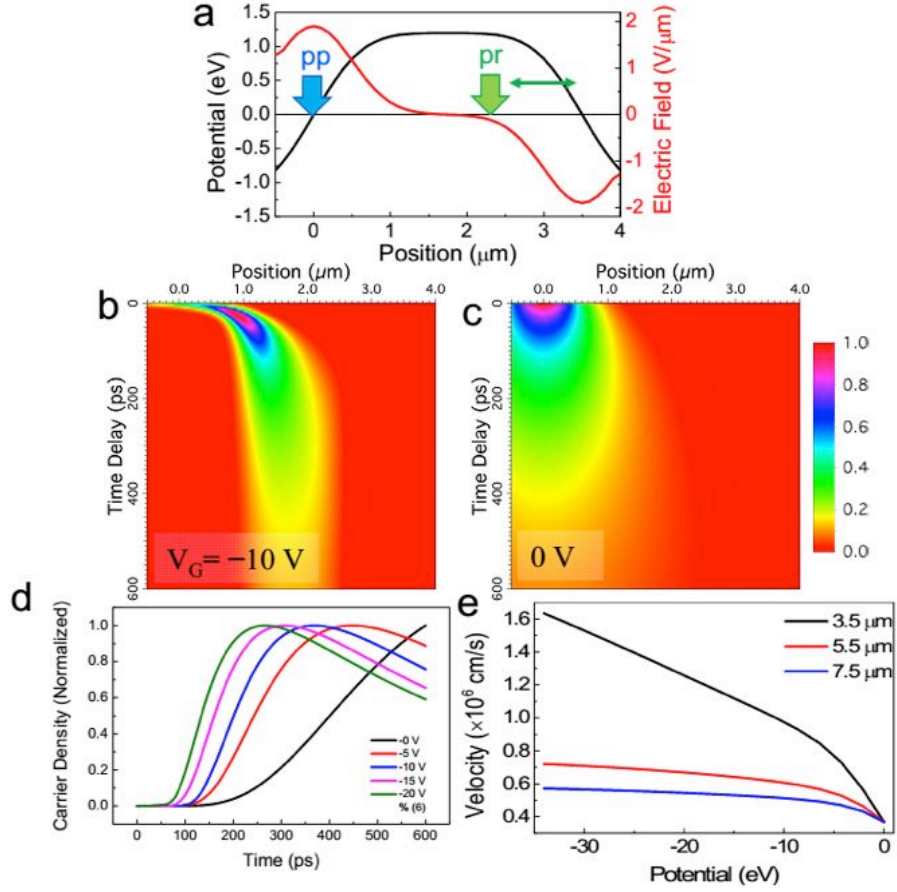

Fig. S11 (a) Position-dependent electrostatic potential and electric field used for simulation. (b) 2D spatio-temporal image of the carrier density when we fix the pump position at $x = 0$. (c) That of (b) with $V_G = 0$ (without the electric fields) (d) Carrier density is plotted as a function of time for the different gate bias voltages measured at $x = 3.5\ \mu$m. (e) Most probable carrier velocity as a function of $V_G$ measured at the three different $d_{pc}$'s.

Our experimental findings were analyzed in term of numerical calculations based on the drift-diffusion model.

$$\frac{\partial n}{\partial t} = D\nabla^2 n + \mu \nabla \cdot (n\vec{E}) - \frac{n}{\tau} + G \qquad (1)$$

where $n$ is the excess carrier density, $D$ is the carrier diffusion coefficient, $\mu$ is the carrier mobility, $\tau$ is the lifetime, $\vec{E}$ is the internal electric field, and $G$ is the generation rate of the carriers excited by the femtosecond laser pulse. The internal electric field is position-dependent, i.e., $\vec{E} = \vec{E}(x)$, with a strong variation near both the metal electrodes. We used a minority carrier hole mobility of 600 cm$^2$/Vs corresponding, to our n-type Si with doping concentration of ~$10^{16}$ /cm$^3$.

We assumed that the charge carriers were generated by the femtosecond laser pulse (at $x = 0$) with a FWHM 600 nm. We used depletion layer width of $L_d = 0.5\mu$m, whereas the carrier life time was assumed to be 500 ps. Fig. S11a shows the potential plots of electric band structure and the corresponding internal electric field of NW used for calculation with a potential of $-1$ eV. The pump and probe pulses located at the maximum electric field position as denoted by and arrows (position of 0 and 3.5 $\mu$m).

Fig. S11b shows an example of the excess charge carrier density calculations plotted as a function of the position (x-axis) and time (y-axis) with a gate voltage of $-10$ V in Fig. 4a in main manuscript (with the gate efficiency parameter $\alpha$). In our calculations, the negative gate voltages represent the p-type operation condition and the gate voltage of 0 V is the condition for the flat band condition. Apparently, the hole carriers generated at $x = 0$ $\mu$m at $t = 0$ ps move rapidly toward the middle of nanowire ($x > 0$) with time. After certain time, however, the carrier movement decreases gradually which is due to the reduced drift process as the carrier moves away from the depletion layer. This is the region in which the carriers are transported through the diffusion process without the drift effect. On the other hand, when the band is flat ($\vec{E} = 0$ V/$\mu$m), the charge carrier transport is entirely dominated by the diffusion process as shown in Fig. S11c. The carriers generated at $t = 0$ diffuse gradually outside the illumination region. This plot can be useful for estimating the arrival time of the carriers generated in the middle of nanowires when there is no built-in electric field as shown previously in Fig. S6c.

The time trace for a given probe position can be extracted from the 2D plot, in other words, along a line for $x = 3.5$ $\mu$m, whose results is plotted in Fig. S11d, for the five different $V_G$'s.

It is obvious that the carriers arrive faster for $V_G = -20$ V (due to the large electric fields) relative to the flat band condition with no built-in electric fields. Finally, in Fig. S11e, we show most probable velocities of carriers as a function of the gate voltage from $-30$ V to $0$ V, for the different probe positions from $3.5$ $\mu$m to $7.5$ $\mu$m. The velocities are very fast when we applied large electric potential (p-type operation) and decreases, until they reach the minimum values at the flat band conditions. Although this results does not reproduce our experimental findings accurately, many of the essential observation has been demonstrated. One of the plausible explanation of the discrepancy is that the energetically hot carriers are generated when we illuminate with the laser energy ($1.5$ eV) higher than Si band-gap ($1.1$ eV).